\documentclass[twocolumn,showpacs,preprintnumbers,prb]{revtex4-1}

\usepackage{eurosym}
\usepackage{graphicx,bm,amsmath,amssymb}

\def\gz{\ifmmode{Z\hskip -4.8pt Z}
    \else{\hbox{$Z\hskip -4.8pt Z$}}\fi}

\newcommand{\be}{\begin{equation}}
\newcommand{\ee}{\end{equation}}
\newcommand{\bea}{\begin{eqnarray}}
\newcommand{\eea}{\end{eqnarray}}

\begin{document}

\title{Charge and spin gaps of the ionic Hubbard model with 
density-dependent hopping}

\author{O. A. Moreno Segura}

\affiliation{Centro At\'{o}mico Bariloche and Instituto Balseiro, 8400 Bariloche, Argentina}

\author{K. Hallberg}
\affiliation{Instituto de Nanociencia y Nanotecnolog\'{\i}a CNEA-CONICET,
Centro At\'{o}mico Bariloche and Instituto Balseiro, 8400 Bariloche, Argentina}

\author{A. A. Aligia}
\affiliation{Instituto de Nanociencia y Nanotecnolog\'{\i}a CNEA-CONICET, GAIDI,
Centro At\'{o}mico Bariloche and Instituto Balseiro, 8400 Bariloche, Argentina}

\begin{abstract}
We calculate the charge gap $\Delta E_C$ and the spin 
gap $\Delta E_S$  of the ionic Hubbard chain including electron-hole symmetric density-dependent hopping. The vanishing of $\Delta E_C$ ($\Delta E_S$) 
signals a quantum critical point (QCP) in the charge 
(spin) sector. Between both critical points, the system is a fully gapped spontaneously dimerized insulator (SDI).
We focus our study in this region.
Including alternation in the hopping, it is possible to perform an adiabatic Thouless pump of one charge per cycle, but with a velocity limited by the size of the gaps.

\end{abstract}



\maketitle

\section{Introduction}

\label{intro}

The ionic Hubbard model (IHM) consists of the usual Hubbard model with on-site Coulomb repulsion $U$ supplemented by an alternating one-particle potential $\Delta$. It has been used to study the neutral-to-ionic transition in organic charge-transfer salts \cite{naga,hub} and the ferroelectric transition \cite{egam}. 
More recent studies have established that the chain has 
three different thermodynamic phases, and different gaps,
correlation functions and other properties have been studied
\cite{sdi,phihm,manih,rihm,exac,sdi3,tinca,abol1,abol2}. 
Here we assume half filling. The unit cell consists of
two sites with on-site energies $\pm \Delta$. Chains with
larger unit cells have also been studied \cite{abn,stenzel}.

The model has three phases, the band insulating (BI), the Mott insulating (MI) and a narrow spontaneously dimerized insulating (SDI) phase in between. An intuitive understanding of the first two phases is provided by the zero-hopping limit, 
in which the occupancies of the different sites are like 
2020... (BI phase) for $\Delta > U/2$ and 1111... (MI phase)
for $\Delta < U/2$. For finite hopping the SDI phase appears
in between, as first shown by bosonization \cite{sdi}, and described in more detail later using an approximate mapping to an SU(3) Heisenberg model \cite{exac,sdi3}.

The phase diagram of the model has been constructed in Ref. 
\onlinecite{phihm} using the method of crossings of excited energy levels (MCEL) based on conformal field theory \cite{nomu,naka,naka1,naka2,som}. For this model (including 
also density-dependent hopping) the method also coincides
with that of jumps of charge and spin Berry phases used in Ref. \onlinecite{phtopo}. The basis of the MCEL is that
in one dimension, the dominant correlations at large distances correspond to the smallest excitation energies. Thus, the crossings of excited levels in appropriate symmetry sectors therefore correspond to phase transitions. 
Lanczos using total wave vector, inversion symmetry \cite{note} and time-reversal symmetry has been used in order to separate the different symmetry sectors, limiting the maximum size to 16 sites. The results were obtained extrapolating to the thermodynamic limit \cite{phihm}. Open-shell boundary conditions (OSBC) were used, which correspond to 
periodic BC for a number of sites $L$ multiple of 4, 
and antiperiodic BC for even $L$ not multiple of 4.

For fixed $U$, 
small $\Delta$, and half-filling as assumed here, 
the system is in the MI phase with zero spin gap. 
Increasing
$\Delta$, at the point $\Delta=\Delta_s$, 
a spin gap $\Delta E_S$
opens signaling the transition to the SDI phase. 
the transition is of the Kosterlitz-Thouless type \cite{sdi}. Although the spin gap is exponentially small near the transition, the MCEL allows to identify it unambiguously 
and accurately from the crossing of the even singlet with lowest energy and the odd triplet of lowest energy
(both states have higher energy than the ground state)-
At $\Delta=\Delta_s$, 
also the spin Berry phase $\gamma_s$ \cite{phihm,phtopo} jumps from
$\pi$ to 0 mod($2\pi$). Further increasing $\Delta$, rather soon, at the point $\Delta=\Delta_c$ a charge transition
from the SDI to the BI phase takes place in which
the charge reorders. At this point, 
there is a crossing of the two singlets of 
lowest energy with opposite parity under inversion. In the BI phase the ground state is the singlet even under inversion, while it is the odd singlet in the other two phases.
All  these states have wave vector 0 for 
$\Delta \neq 0$. In turn, this crossing leads to a jump 
in the charge Berry phase $\gamma_c$ from $\pi$ to 0 mod($2\pi$). 
As explained above, for $\Delta=\Delta_c$ and using OSBC, the charge gap $\Delta E_C$ defined as the absolute value of difference 
in energy between the ground state and the first excited 
state at half filling (in other works called exciton gap \cite{manih} or internal gap \cite{eric}) vanishes 
at the charge transition.

Changes in $\gamma_c$ are proportional to
changes in the polarization. Actually, calculations of 
the charge Berry phase form the basis of the modern 
theory of polarization 
\cite{pola1,pola2,pola3,bradlyn,om,resor,oc,song,zup}.
A jump in $\pi$ in $\gamma_c$ is consistent with a displacement of an electronic charge per unit cell
in half a unit cell (to the next site) on average. This is the change of polarization that corresponds to the change in site occupancies from 1111.. to 2020...
The IHM in a ring has inversion symmetry with center at any site \cite{note}, and as a consequence $\gamma_c$ and $\gamma_s$ 
can only be 0 or $\pi$ mod($2\pi$). In other words,
they are $Z_2$ topological numbers protected by inversion
symmetry \cite{zup}.

\begin{figure}[hb]
\begin{center}
\includegraphics[width=7cm]{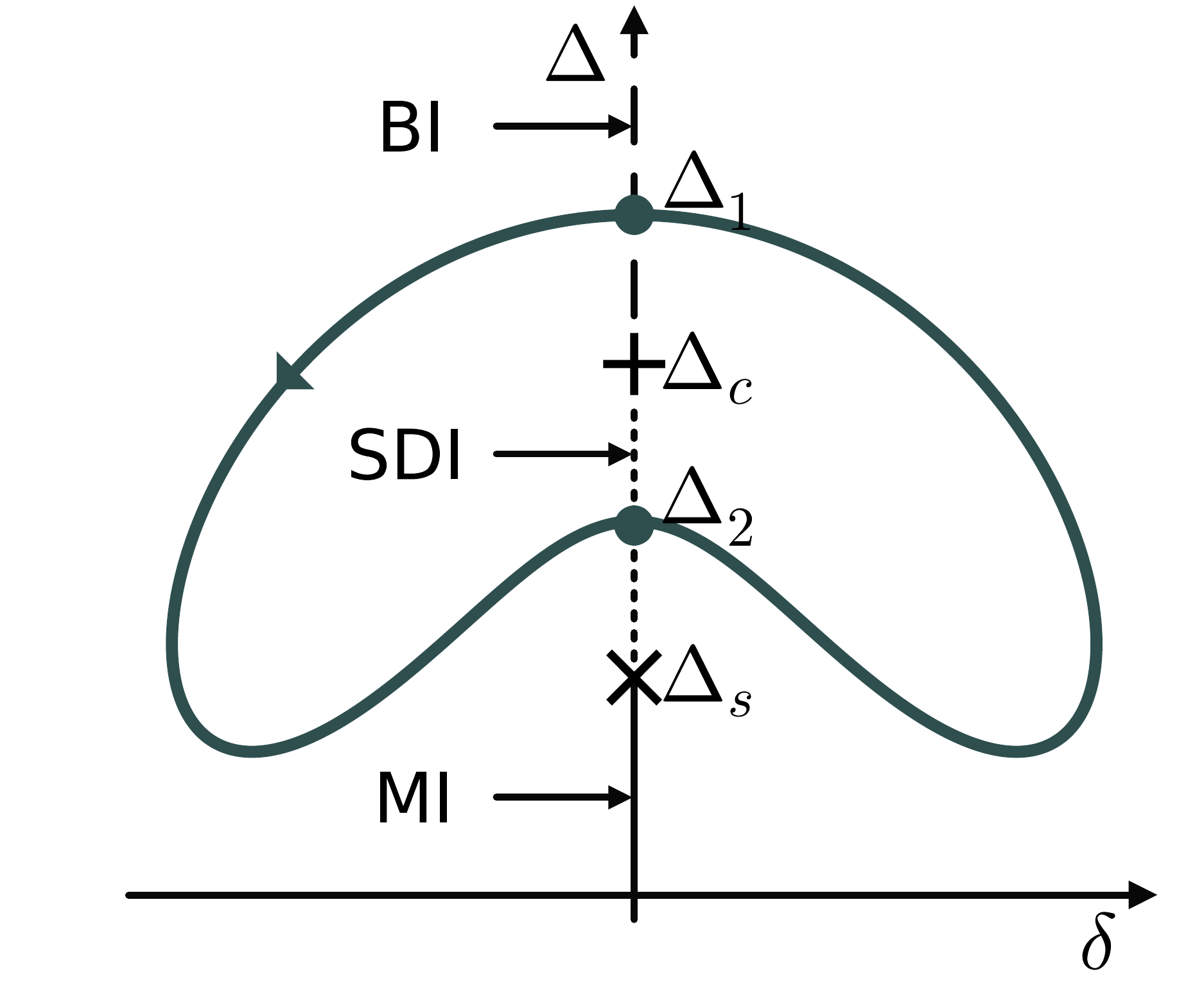}
\caption{Schematic representation of the pump trajectory. Dashed, dotted, and solid lines indicate the BI, SDI and MI phases of the IHM (at $\delta = 0$), respectively.}
\label{cycle}
\end{center}
\end{figure}

If a modulation of the hopping $\delta$ is introduced,
the inversion symmetry is lost, and $\gamma_c$ can change continuously. This permits to transfer one charge to the next
unit cell in a Thouless pump cycle in the $(\Delta,\delta)$
plane (see Fig. \ref{cycle}). This can be understood as follows. Starting at a point
$(\Delta_1,0)$ with  $\Delta_1 > \Delta_c$, $\gamma_c=0$.
Then introducing a finite $\delta$, with the appropriate sign, $\gamma_c$ increases continuously with increasing $|\delta|$.
Decreasing $\Delta$ to a value $\Delta_2 < \Delta_c$ and returning $\delta$ to zero, the point $(\Delta_2,0)$ 
is reached where $\gamma_c=\pi$. Continuing the cycle with the opposite sign of $\delta$, $\gamma_c$ continues to increase 
and reaches the value $\gamma_c= 2 \pi$ at the end of the cycle at $(\Delta_1,0)$. This corresponds to the displacement 
of one unit charge by one unit cell according to the modern theory of polarization. The values of the Berry phases 
in the cycle and time-dependent calculations of the charge transferred have been presented in Ref. \onlinecite{eric}. Moreover, this pumping procedure has been realized 
experimentally recently \cite{expe}, allowing to study the effects of interactions in the field of quantized
topological charge pumping in driven systems,
which is of great interest in the last years \cite{citro,konrad}.

A problem with the pumping cycle mentioned above is that 
it usually crosses the MI segment between the points 
$(\Delta_s,0)$ and $(-\Delta_s,0)$ at which the spin gap vanishes. Since unavoidably this segment is traversed at 
a finite speed, spin excitations are created, leading 
to the loss of adiabatic quantized pumping \cite{eric,expe}
(we note that introducing $\delta$ in the MI phase, a spin 
gap $\Delta E_S$ opens proportional to $|\delta|^{2/3}$ for small 
$\delta$ \cite{eric}). To avoid this problem, one might choose
the crossing point $\Delta_2$ inside the SDI phase, that is
$\Delta_s < \Delta_2 < \Delta_c$ (as it is shown in Fig. \ref{cycle}), and then the system 
is fully gapped in the whole trajectory. However, 
at $\Delta=\Delta_2$ both gaps $\Delta E_C$ and $\Delta E_S$ 
are small,  and their magnitude is not known.
Previous calculations of $\Delta E_S$ were affected by strong
finite-size effects in the SDI region and were limited 
to very large values of $U$ \cite{manih}. On the other hand,
it has been recently shown that the SDI phase is enlarged at 
small values of $U$ if density dependent hopping is 
introduced \cite{tab}. A density-dependent hopping can 
be experimentally engineered by near-resonant Floquet modulation \cite{ma,meine,go1,messer,goerg}.

In this work, we calculate both gaps, $\Delta E_C$ and 
$\Delta E_S$ inside and near the SDI phase, and explore
the optimum value of $\Delta_2$ for which the smallest gap is maximum. We use density-matrix renormalization group (DMRG) \cite{S.white,karen_rev,Uli,Uli2,banuls}, as described in Section \ref{methods}. We find that to calculate
$\Delta E_S$ open BC are more convenient, while to calculate
$\Delta E_C$, a ring with OSBC leads to the optimum results, improving previous estimates and allowing to calculate the charge gap within the SDI phase with unprecedented accuracy.

The paper is organized as follows. In Section \ref{model} we briefly explain the model. In Section \ref{methods}, we describe  the  methods used to calculate the gaps.
The results are presented in Section \ref{res}. Section \ref{sum} contains a summary and discussion.

\section{Model}
\label{model}

The model we study here is the IHM with density-dependent hopping (DDH). It is the version without alternation
of the hopping ($\delta=0$) of the interacting Rice-Mele 
model \cite{rice} including DDH. 
Because of its relevance for quantized charge 
pumping, we describe the full Hamiltonian 
including also $\delta$ below

\begin{eqnarray}
H &=&\sum_{j \sigma} \left[ -1+\delta \;(-1)^{j}\right]
 \left( c_{j\sigma }^{\dagger }c_{j+1\sigma
}+\text{H.c.}\right)   \notag \\
&&\times [t_{AA}(1-n_{j\bar{\sigma}})(1-n_{j+1\bar{\sigma}}) 
+t_{BB}n_{j\bar{\sigma}}n_{j+1\bar{\sigma}}
\notag \\
&&+t_{AB}(n_{j\bar{\sigma}}
+n_{j+1\bar{\sigma}}-2n_{j\bar{\sigma}}n_{j+1\bar{\sigma}})]  \notag \\
&&+\Delta \sum_{j \sigma}(-1)^{j}n_{j\sigma }
+U\sum_j n_{j\uparrow }n_{j\downarrow }.  \label{hirm}
\end{eqnarray}
The first term is the DDH, which is alternating 
for $\delta \neq 0$. The amplitudes $t_{AA}$, $t_{AB}$ and $t_{BB}$ correspond to hopping of a particle with a given spin, when the total occupancy of both sites for particles with the opposite spin is 0, 1 and 2 respectively. In the following we assume the 
electron-hole symmetric case $t_{BB}=t_{AA}$, which is the 
one implemented experimentally with cold atoms \cite{ma,meine,go1,messer,goerg}. 
$\Delta$ is the alternating on-site energy and $U$ is the 
on-site Coulomb repulsion. 

The model with $\Delta=\delta=0$ has been derived and studied in two dimensions as an effective model for cuprate superconductors \cite{fedro,brt,hir2,lili}. In one 
dimension also superconductivity is favored for some parameters \cite{inco,dobry,ari,chen,jiang}. Our interest in DDH here is that for $t_{AB}$ larger than the other two, 
the fully gapped SDI phase is favored \cite{tab,naka2,phtopo,jaka,bosolili}. This is important 
for fully adiabatic quantized charge pumping of one charge.
So far, charge pumping has been studied in the interacting
Rice-Mele model in absence of DDH ($t_{AA}=t_{AB}=t_{BB}$)
\cite{eric,expe,konrad,nakag}.

\section{Methods}
\label{methods}
To perform the energy level calculations, we have used the DMRG method with a code that relies on the ITensors library for Julia \cite{itensors}. Conveniently setting $S_z$ sectors, we have calculated ground and excited states with a fixed bond dimension of 900. The truncation error is, in the worst case, on the order of $10^{-6}$ for periodic BC (PBC), and 
$10^{-10}$ for open BC (OBC).

In general, it is convenient to use OBC rather than PBC, because the entanglement is lower in the former case, leading to more accurate results in less amount of time. In turn, this fact
permits to reach larger systems. This is particularly important for the spin gap $\Delta E_S$, because even in regions of parameters for which $\Delta E_S=0$ in the 
thermodynamic limit, it is finite for finite systems,
scaling as $1/L$ for increasing system size $L$ \cite{nomu}.
We have calculated the spin gap by extrapolating the results 
for different system sizes using
a quadratic function in $1/L$. The calculations were done for systems between $L = 40$ and $L=100$, except in the case of $U=10$, where we have used sizes up to $L=64$.

\begin{figure}[ht]
\begin{center}
\includegraphics[width=7cm]{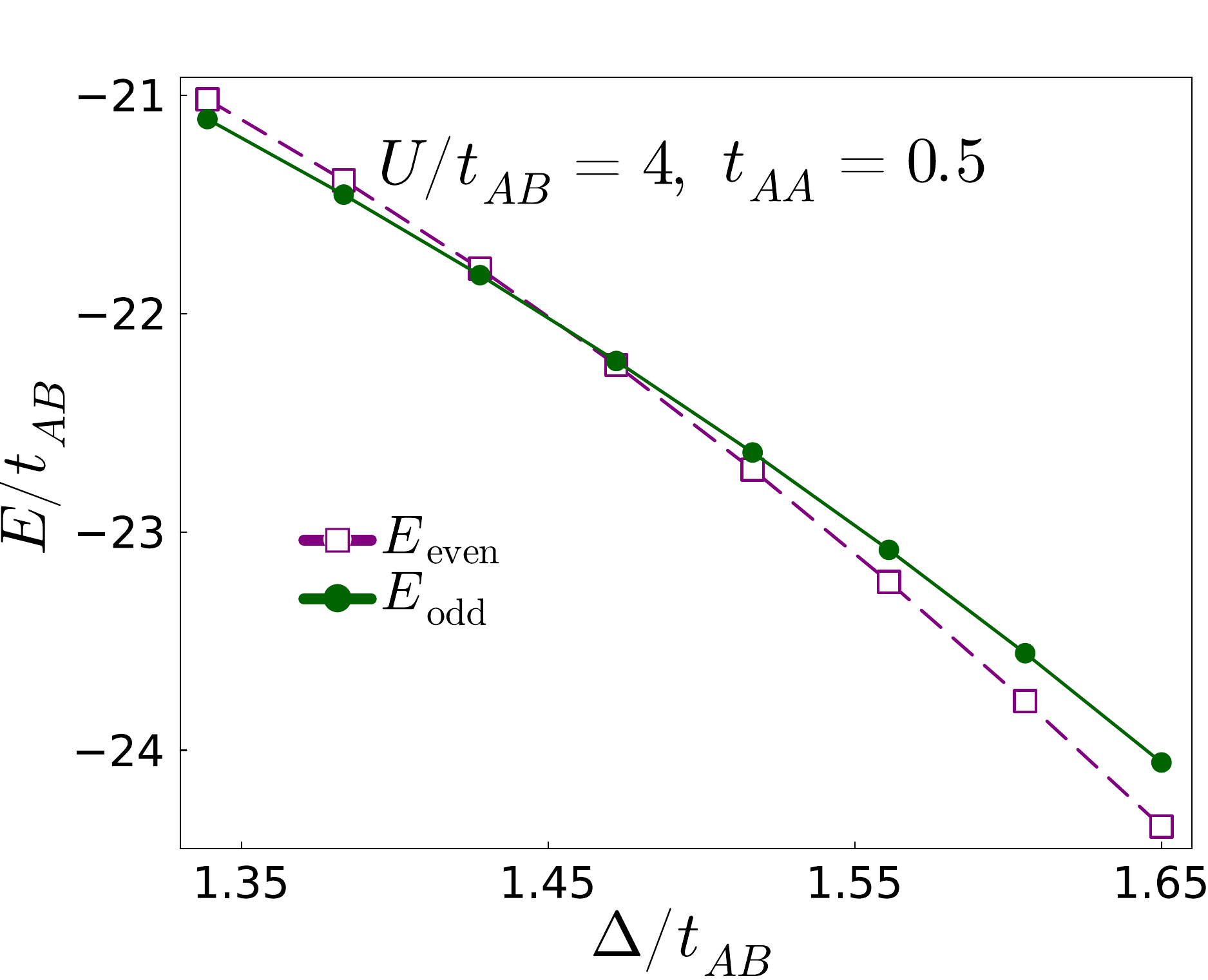}
\caption{(Color online) Ground state and 
first excited state as a function of $\Delta$ 
for 32 sites and PBC.}
\label{cros}
\end{center}
\end{figure}

For the charge gap $\Delta E_C$, which is the difference
of energies between the first excited state and the ground state in the singlet sector, the situation is different. 
For OBC we find similar difficulties as those found 
before \cite{manih} for calculating the gap in the SDI 
phase and particularly near the transition to the BI phase,
where it should vanish in the thermodynamic limit. The reason is the following. As it is clear using the MCEL method
mentioned in Section \ref{intro}, the ground state and the first excited state have opposite parity under inversion, 
being the even state the one of lowest energy in the BI 
phase, and both states cross at the BI-SDI transition.
For a chain with OBC and an integer number of unit cells,
the inversion symmetry is lost, the crossing becomes an  anticrossing, and extrapolation to the thermodynamic limit 
becomes problematic. Therefore, we change the method 
using a ring with OSBC as described below.

\begin{figure}[ht]
\begin{center}
\includegraphics[width=7cm]{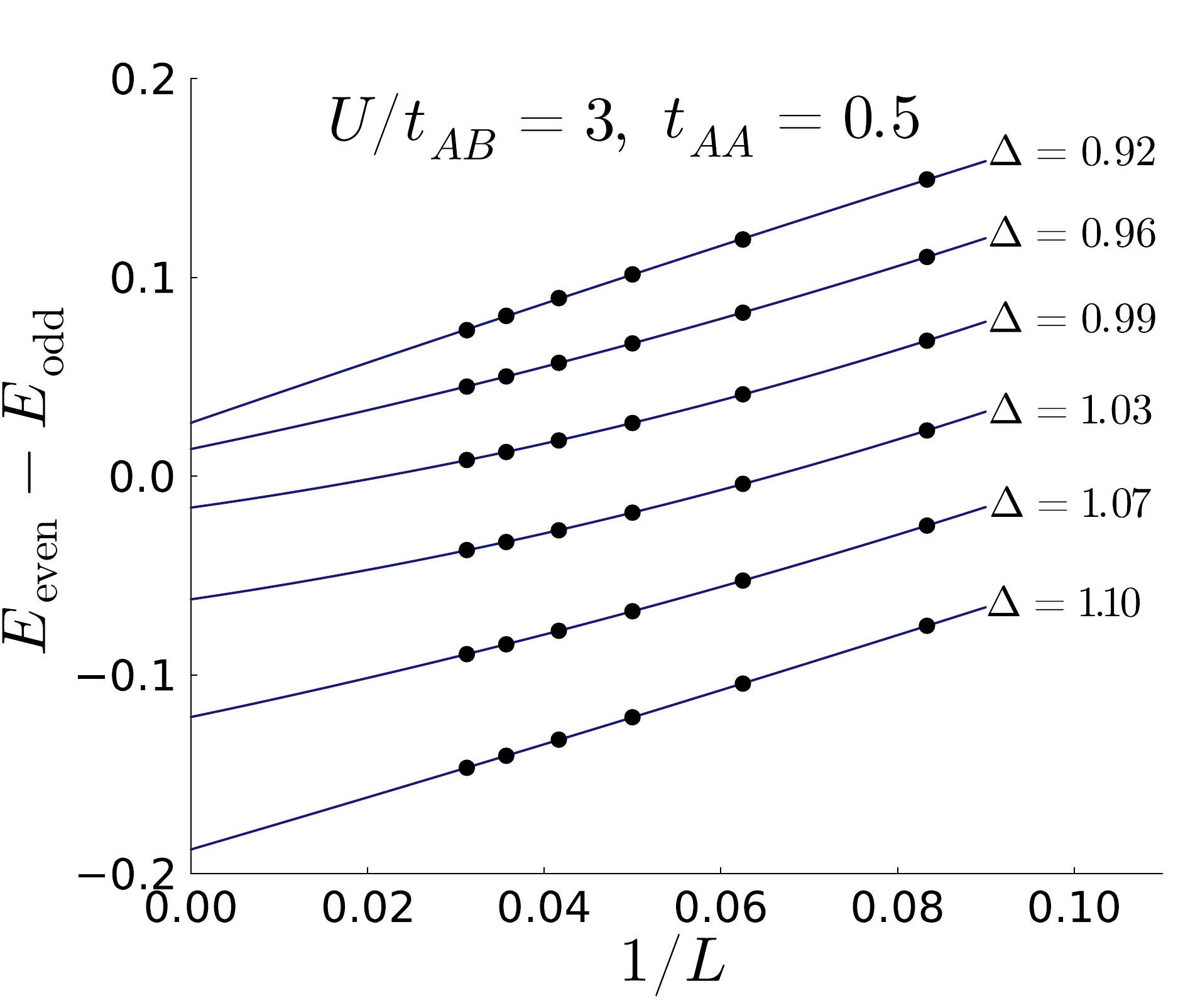}
\caption{Difference of energy between
the even and odd states of lowest energy as a function of
the inverse of the systems size $L$ for all $L$ multiple of 4
in the range $12 \leq L \leq 32$ with PBC for several values of $\Delta$. The transition is calculated to be at 
$\Delta_c=0.978$.}
\label{extrap}
\end{center}
\end{figure}

The Lanczos method used in the MCEL has divided the Hilbert space in different symmetry sectors, but the method is limited to 16 sites at half filling \cite{phihm,naka2,tab}. Our method allows us to use larger system sizes, 
but we do not have access to the different symmetry sectors.
In any case, just plotting the energy of the ground state
and first excited state as a function of $\Delta$ in a ring, 
both energy levels and the crossing can be clearly identified.
This is illustrated in Fig. \ref{cros} for a typical case.
$t_{AB}=1$ is chosen as the unit of energy. 
We find that extrapolating the energies of $L$ multiple
of 4 for a ring with PBC (which coincide with OSBC for the chosen $L$) between $L=12$ and $L=32$ using
a quadratic function in $1/L$, an accurate and reliable result
for $\Delta E_C$ in the thermodynamic limit is obtained.
An example of the extrapolation is presented in Fig. \ref{extrap}

Noting that the slopes of the gap are different at both sides
of the transition, we find that the difference between
odd and even states can be well fitted by the following function with three parameters

\begin{equation}
E_{\mathrm{odd}}-E_{\mathrm{even}} = (\Delta -\Delta_c)
\left[A+\text{tanh}\left(\frac{\Delta -\Delta_c}{B} \right) \right].  
\label{dif}
\end{equation}
Examples will be shown in the next Section.

Comparing with previous results using the MCEL in smaller systems \cite{tab}, we have also found that using OSBC, 
the crossing
between the first excited state in the sector with total spin projection $S_z=0$ 
(corresponding to the even singlet \cite{tab})
and the lowest-energy state in the sector with  $S_z=1$
(an odd triplet \cite{tab}) corresponds to the crossing 
at $\Delta=\Delta_s$
that signals the opening of the spin gap $\Delta E_S$,
and the SDI-MI transition, as explained in Section \ref{intro}.

Therefore, our methods might be also used to improve the accuracy of phase diagrams calculated with the MCEL, extending the results to larger systems.

We have found empirically, that sufficiently far from 
$\Delta_s$ and in the SDI phase, or in the BI phase near
the SDI-BI transition at $\Delta=\Delta_c$, the dependence on
$\Delta$ of the spin gap is well described by the expression

\begin{equation}
\Delta E_S=A_s\text{exp} 
\left[ B_s \left( \Delta -C_s \right) \right].  
\label{sgap}
\end{equation}

\section{Results}

\label{res} 

\begin{figure}[hb]
\begin{center}
\includegraphics*[width=0.8\columnwidth]{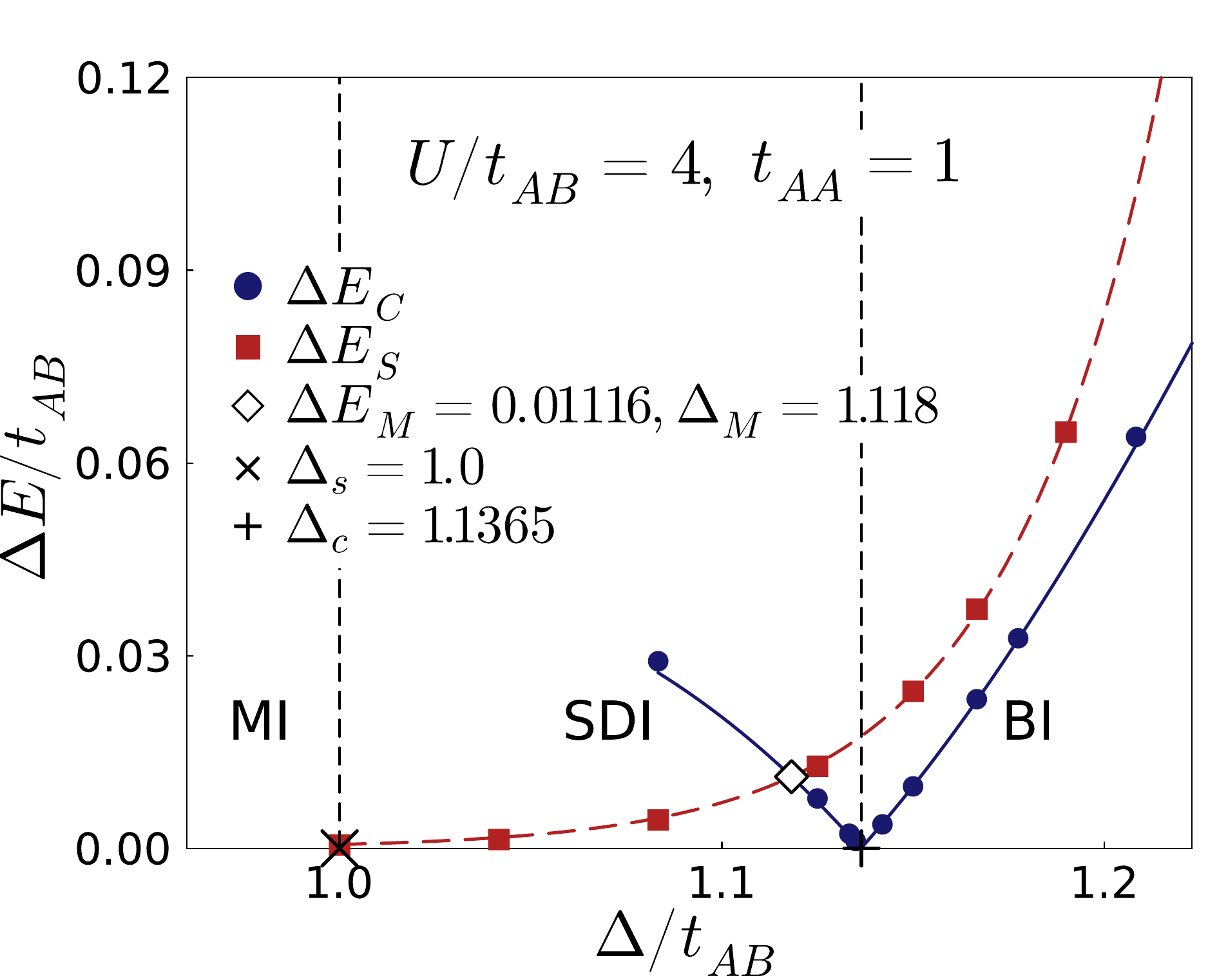}\\
\includegraphics*[width=0.8\columnwidth]{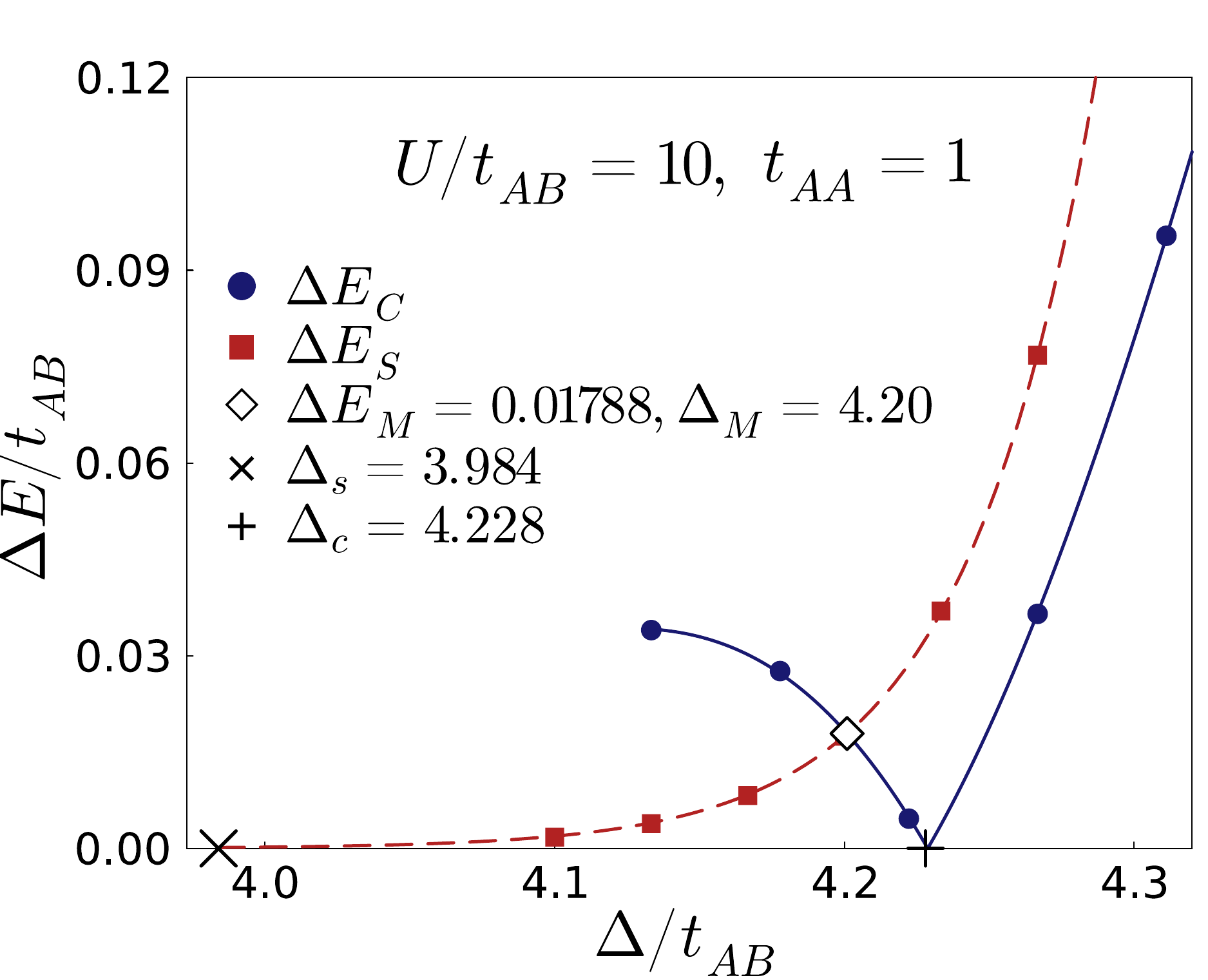}
\end{center}
\caption{(Color online) Charge gap (blue circles) 
and spin gap (red squares) 
as a function of $\Delta$ for two values of $U$ 
and $t_{AA}=t_{BB}=t_{AB}=1$. Blue solid [red dashed] line is a fit using Eq. (\ref{dif}) [Eq. (\ref{sgap})].
Vertical lines in the top figure separate the different phases of the IHM.}
\label{figsch}
\end{figure}

In Fig. \ref{figsch}, we show the gaps for the model 
without DDH and two values of $U$. The maximum difference between any excited state and the ground state 
in the SDI phase is denoted by 
$\Delta E_M$. This value is obtained at the crossing
between both studied gaps.
The value of $\Delta$ at this crossing is denoted as 
$\Delta_M$. We take $t_{AB}=1$ as the unit of energy.

From the figure, one can see that 
the spread of the SDI phase $\Delta_c-\Delta_s$ 
and also $\Delta E_M$ are larger 
for larger values of $U$ than for moderate ones.
The former fact is in agreement with calculations of
the phase diagram using up to 16 sites \cite{phihm},
although $\Delta_c-\Delta_s$ is a little bit smaller
in our case. Our values should be more accurate since 
we have calculated $\Delta_c$ and $\Delta_s$ using up to 32 and 28 sites, respectively.

\begin{figure}[th]
\begin{center}
\includegraphics*[width=0.8\columnwidth]{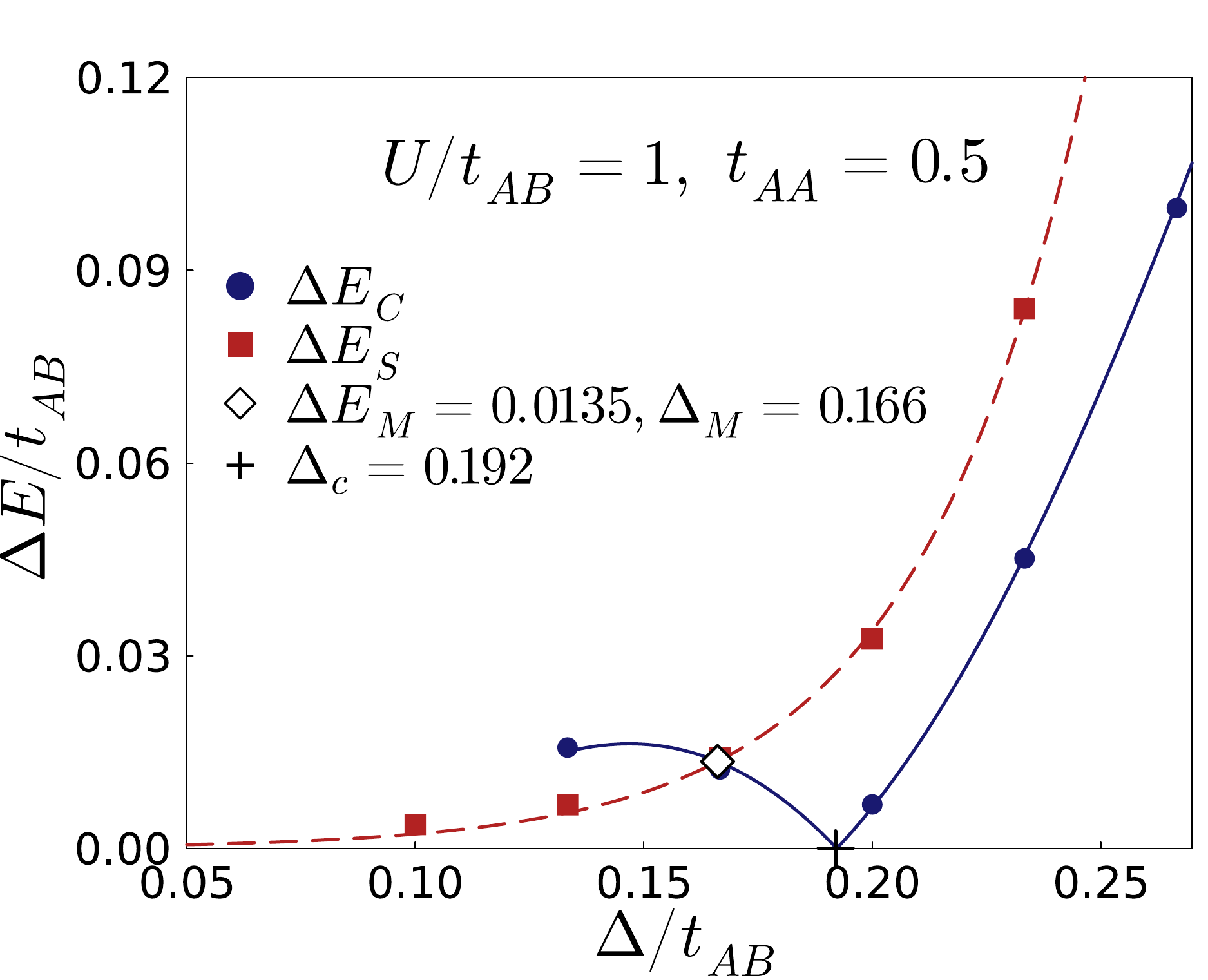}\\
\includegraphics*[width=0.8\columnwidth]{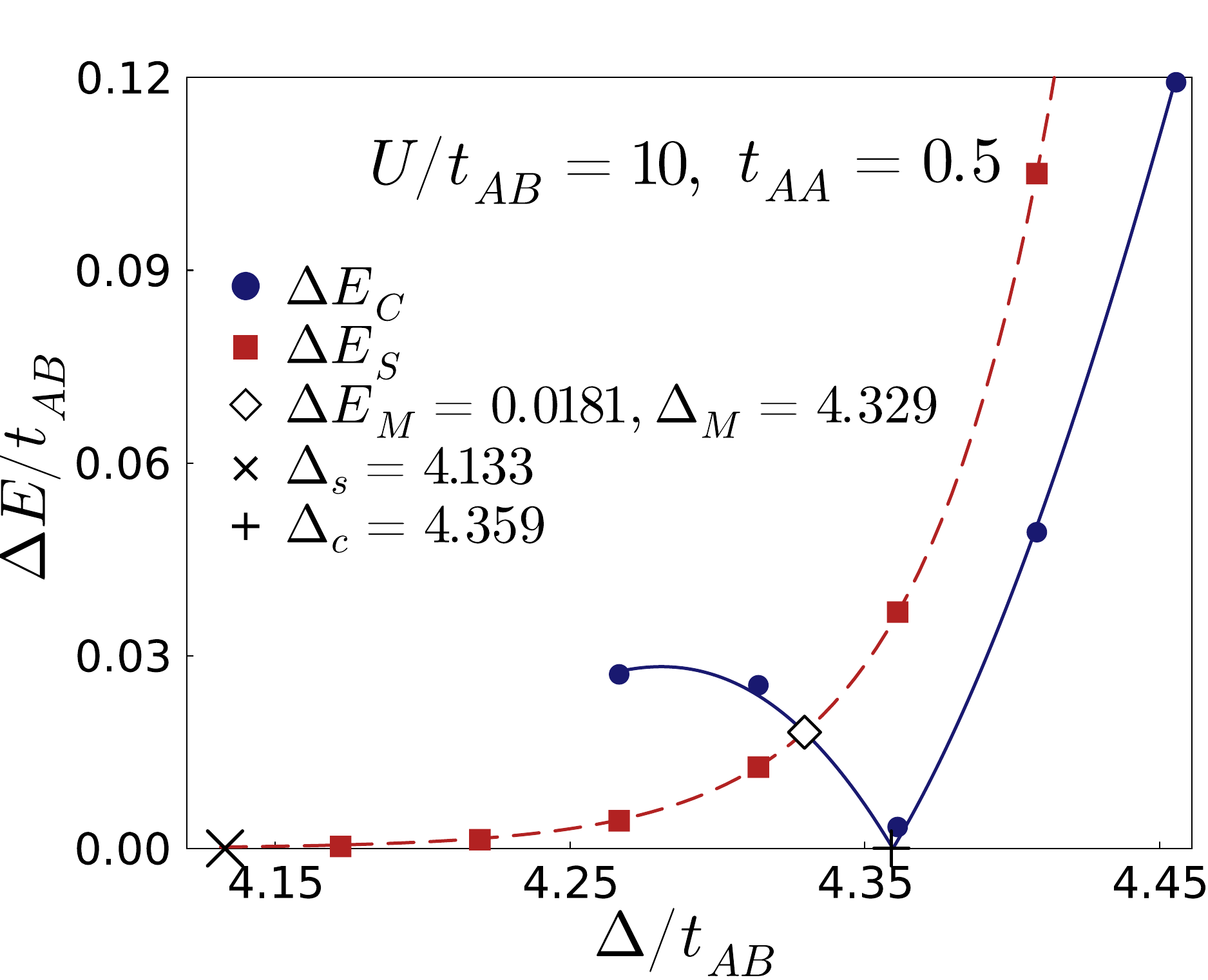}
\end{center}
\caption{(Color online) Same as Fig. \ref{figsch} for $t_{AA}=t_{BB}=0.5$ and $t_{AB}=1$.}
\label{figche}
\end{figure}

\begin{figure}[tbp]
\begin{center}
\includegraphics*[width=0.8\columnwidth]{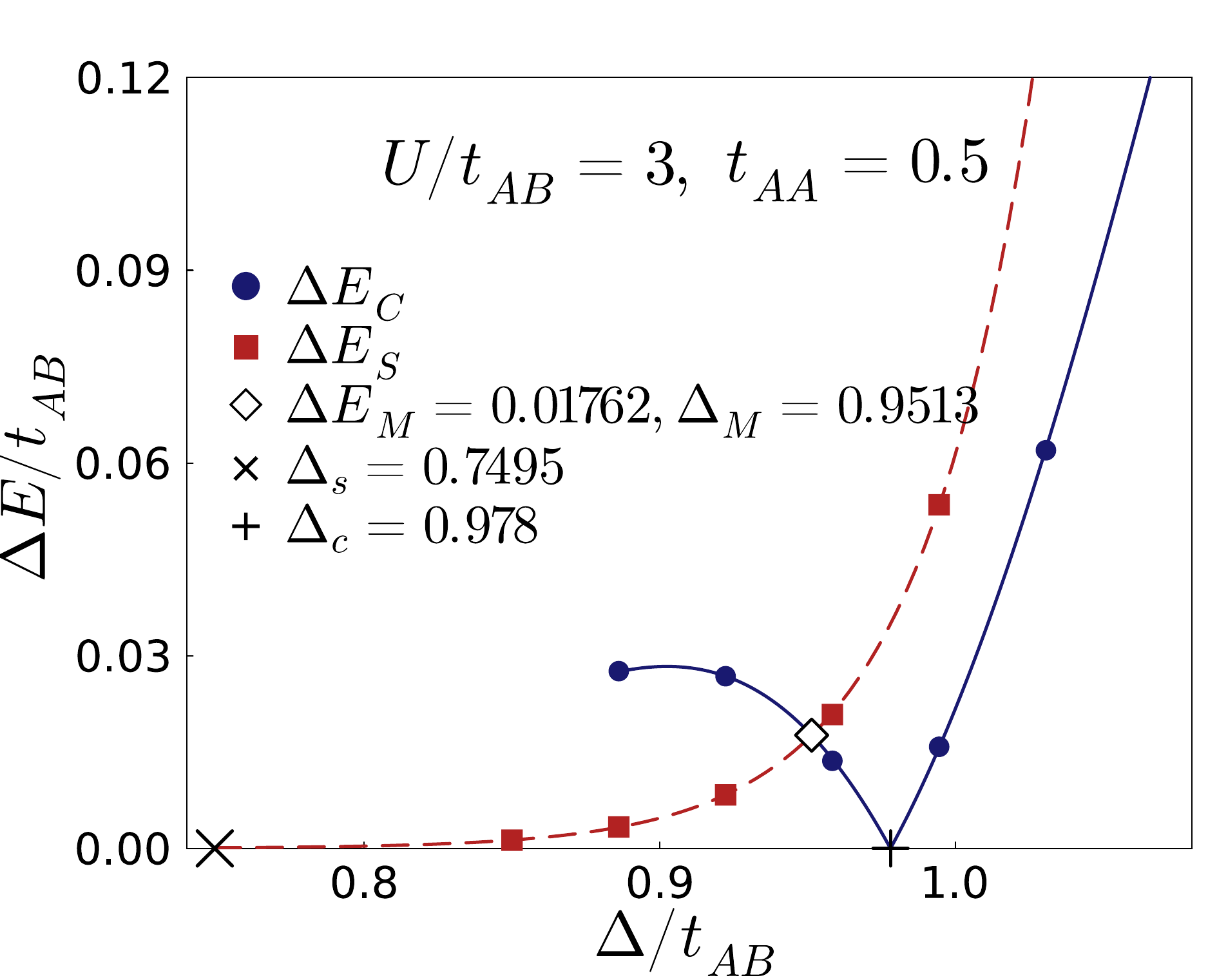}\\
\includegraphics*[width=0.8\columnwidth]{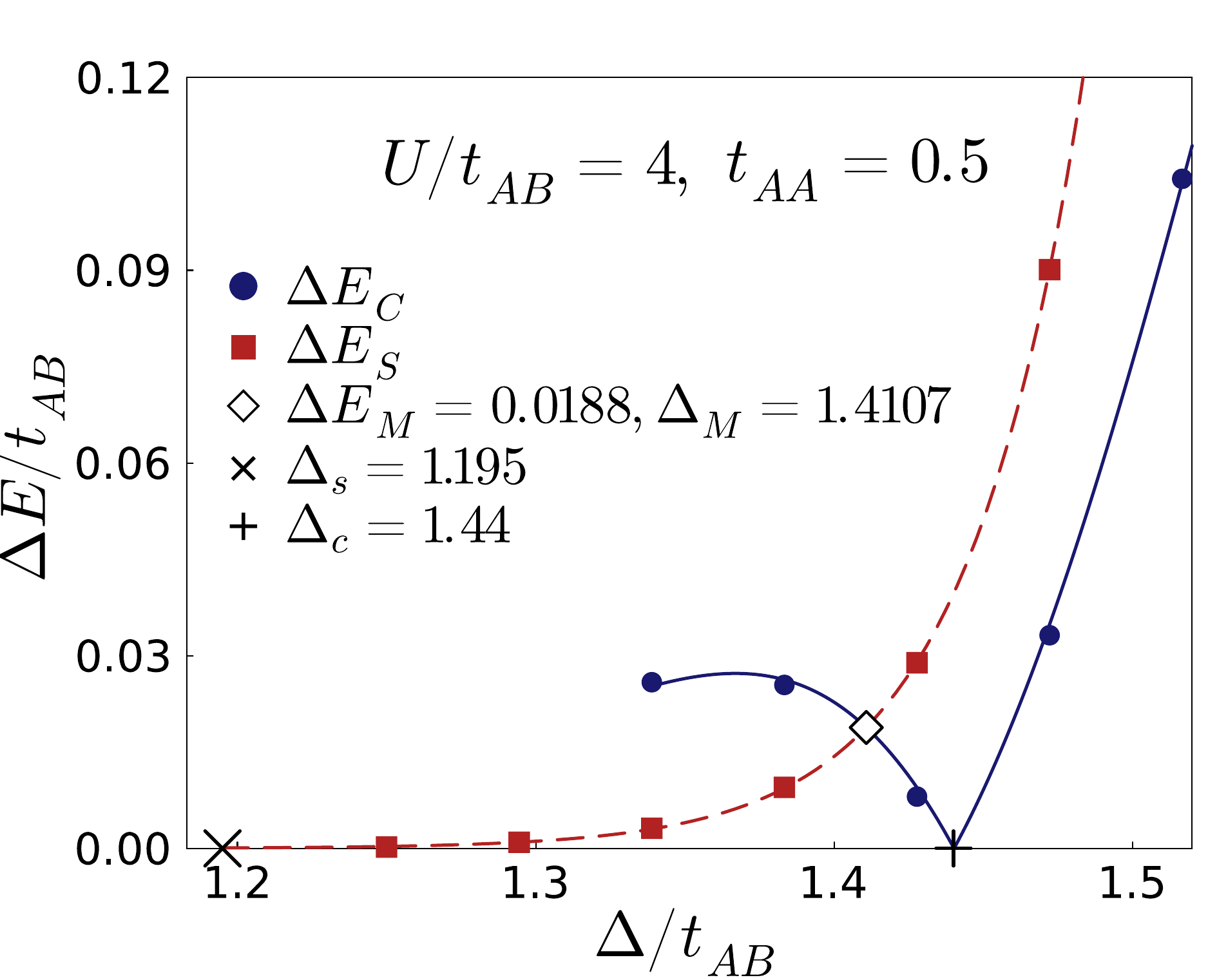}\\
\includegraphics*[width=0.8\columnwidth]{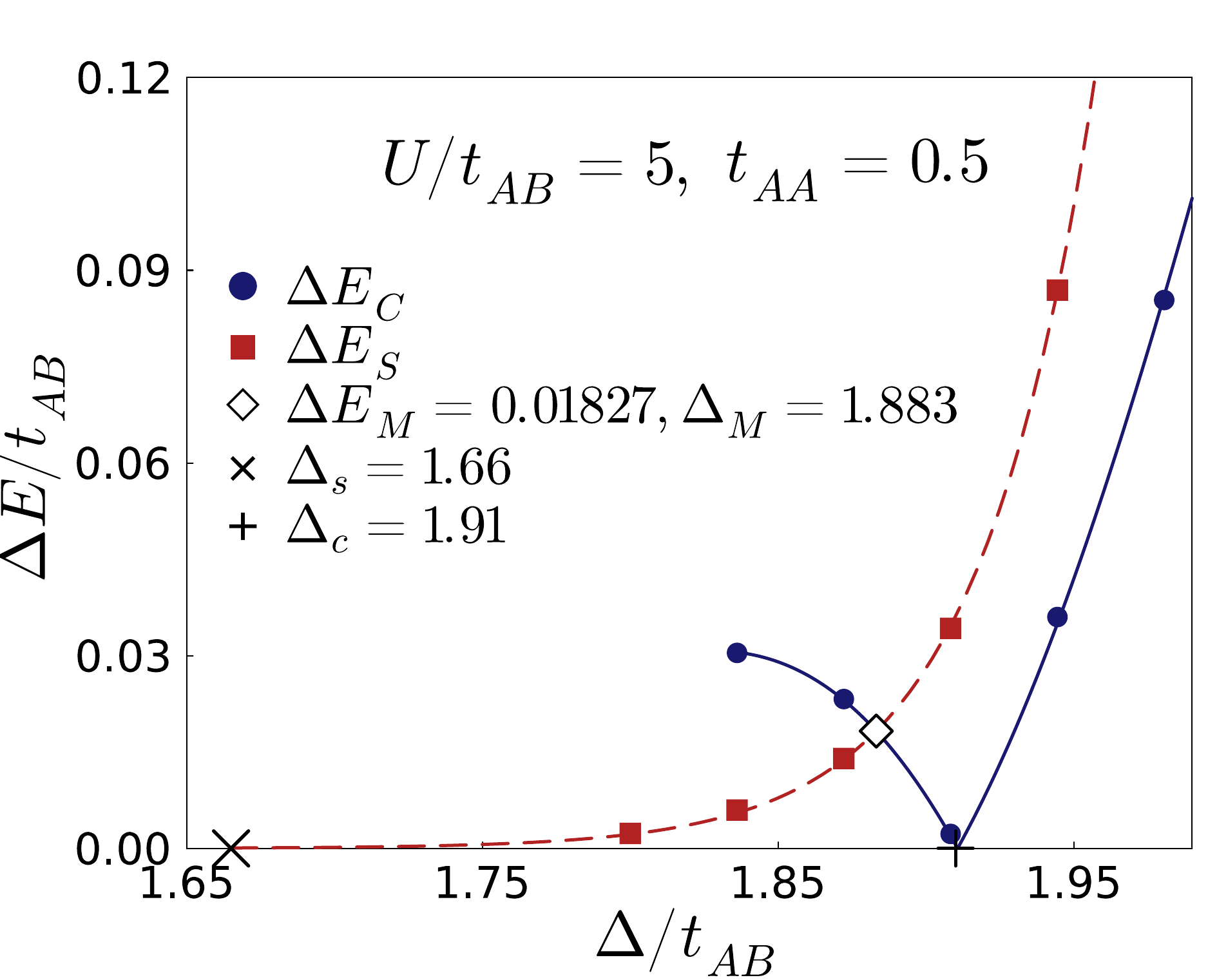}
\end{center}
\caption{(Color online) Same as Fig. \ref{figche} for 
intermediate values of $U$.}
\label{figch}
\end{figure}

In Fig. \ref{figche} we analyze the effect of DDH,
decreasing $t_{AA}=t_{BB}$ to half the value of $t_{AB}=1$,
for two 
extreme values of $U$, leaving the intermediate values of $U$ 
for Fig. \ref{figch}. For $U=10$, the maximum value of the 
gap $\Delta E_M$ \textit{increases} slightly. This effect is rather surprising, because one naively expects that reducing the average value of the hopping, both $\Delta E_M$ and the 
amplitude of the SDI phase should decrease. Therefore, the 
effect of introducing DDH overcomes the effect of reducing the average hopping regarding $\Delta E_M$. Instead, the amplitude of the SDI phase $\Delta_c-\Delta_s$ decreases slightly.

As discussed earlier \cite{tab}, for small values of $U$,
the amplitude of the SDI phase increases strongly, since it
continues to exist even for $\Delta=0$. However, the magnitude of the maximum gap $\Delta E_M$ is 
reduced by 25\% when $U$ is reduced from
10 to 1 in units of $t_{AB}$.

In order to look for the largest possible value of
$\Delta E_M$ in presence of DDH, we have calculated the gaps for intermediate values of $U$. 
The result is shown in Fig. \ref{figch}.
While qualitatively, the results for $U=3$, 4 and 5 are similar, the maximum gap $\Delta E_M=0.0188$ is obtained for 
$U=4$.

\section{Summary and discussion}

\label{sum} 

We have calculated the charge and spin gaps of the 
spontaneously dimerized insulating (SDI) phase of the ionic 
Hubbard model, including electron-hole symmetric density-dependent hopping.
We have developed a new method using DMRG to calculate the 
charge gap, which presents advantages with respect to previously used ones, leading to substantially more accurate values. In addition, phase diagrams
constructed by the method of crossing of energy levels 
might be calculated more accurately than using only
Lanczos methods, if they are combined with DMRG (the 
former can be used to identify the symmetry sectors).

The results might be useful to present experiments with
cold atoms in which quantized Thouless pumping of 
one charge is observed, when a pump cycle in the 
two-dimensional space $(\Delta,\delta)$ enclosing
the point $(\Delta_c,0)$ is performed in a 
realization of the interacting Rice-Mele model 
[Eq. (\ref{hirm})], where $\Delta_c$ is the value of
$\Delta$ at the transition between the SDI and the 
band insulating (BI) phase. A fully adiabatic pump 
is possible if the Mott insulating (MI) phase is avoided.
This phase lies
at the segment between the points 
$(\Delta_s,0)$ and $(-\Delta_s,0)$, where $\pm \Delta_s$ 
are the points of the MI-SDI phase transitions. 
For this purpose, the SDI phase should be traversed. 

Fixing $t_{AB}=1$ we find that the maximum gap inside the 
SDI phase is about 0.019. This is a rather small value,
which by simple estimates seems 
to require a velocity about 10 times smaller 
than that used in available experiments \cite{expe} 
to guarantee adiabatic
pumping in crossing the point $(\Delta_M,0)$. However introducing $\delta$ the gap increases quickly 
(as $|\delta|^{2/3}$ in the MI phase). A time-dependent calculation, possibly decreasing the velocity near
$(\Delta_M,0)$ would be useful to check this procedure.

The effect of density-dependent hopping, reducing  
$t_{AA}=t_{BB}$ and keeping $t_{AB}=1$ is moderate in increasing the gap, although it is important if the average hopping is kept at the same value. Its main effect is that for small $U$, the extension of the fully gapped SDI phase 
is strongly increased.

\section*{Acknowledgments}

AAA (KH) acknowledges financial support provided by PICT 2017-2726 and PICT 2018-01546 (PICT 2018-01546) of the ANPCyT, Argentina. KH acknowledges support from ICTP through the Associates Programs.


\begin{thebibliography}{99}
\bibitem{note} The inversion symmetry in a ring of $L$ sites with center at site $i$ is defined as the operation which 
interchages the sites $i-j$ and $i+j$ mod($L$) for all $j$.
\end{thebibliography}

\end{document}